\documentclass[conference]{IEEEtran}
\usepackage{graphicx}
\usepackage{amsmath}
\usepackage{cite}
\usepackage{url}
\usepackage{booktabs}
\usepackage{multirow}
\usepackage{array}

\title{Early-Stage IoT Device Identification Using Passive Network Traffic Analysis}

\author{
    \IEEEauthorblockN{
    Alex Ciechonski\IEEEauthorrefmark{1},
    Fabio Palmese\IEEEauthorrefmark{2},
    Alessandro E. C. Redondi\IEEEauthorrefmark{2},
    Anna Maria Mandalari\IEEEauthorrefmark{1}}
    
\IEEEauthorblockA{\IEEEauthorrefmark{1}\{alex.ciechonski.22, a.mandalari\}@ucl.ac.uk, \IEEEauthorrefmark{2}\{fabio.palmese, alessandroenrico.redondi\}@polimi.it\\
\IEEEauthorrefmark{1}University College London, \IEEEauthorrefmark{2}Politecnico di Milano} }

\begin{document}

\IEEEoverridecommandlockouts
\IEEEpubid{\makebox[\columnwidth]{978-1-5386-5541-2/18/\$31.00~\copyright2018 IEEE \hfill}
\hspace{\columnsep}\makebox[\columnwidth]{ }}

\maketitle

\IEEEpubidadjcol

\begin{abstract}
The rapid proliferation of Internet of Things (IoT) devices introduces significant security challenges due to limited visibility and weak device-level guarantees. Accurate and timely identification of devices is essential for enforcing network policies and detecting unauthorised hardware, yet existing approaches often rely on long-term traffic observation, payload inspection, or infrastructure-dependent features.
In this paper, we investigate whether IoT devices can be reliably identified during the early stages of network attachment using only passive traffic analysis. We propose a lightweight approach based on flow-level features extracted from metadata, avoiding payload inspection and active probing.
Through systematic evaluation across multiple observation windows, we show that device-specific signatures emerge within the first few seconds of communication, enabling high-accuracy identification (up to 99\%) across 37 IoT devices. Notably, extending the observation window does not consistently improve performance and may slightly degrade accuracy, indicating that the most discriminative behaviour occurs during initial device startup.
These findings demonstrate the feasibility of fast, privacy-preserving IoT device identification at the network edge, supporting real-time enforcement, device inventory, and anomaly detection in practical deployments.
\end{abstract}


\section{Introduction}
The exponential growth of the Internet of Things (IoT) has vastly expanded the attack surface of modern networks. Often deployed with minimal security and limited visibility, these devices are frequent targets for compromise. A core challenge for network operators is maintaining an accurate device inventory; without reliable identification, enforcing access control and detecting unauthorised hardware becomes nearly impossible.
While prior work has demonstrated high identification accuracy using long-term traffic aggregation or domain-based features, the effectiveness of early-stage, metadata-only identification remains underexplored. In particular, it is unclear whether reliable device identification can be achieved within the first seconds of network attachment, without relying on payload inspection or infrastructure-dependent signals.
In this paper, we investigate identifying IoT devices using only the traffic observed during their initial network attachment phase. During this window, devices typically perform deterministic actions, such as DNS resolution and cloud synchronization, that reflect unique vendor implementation choices. We propose a passive, flow-level identification approach that avoids payload inspection, making it suitable for privacy-sensitive environments. Unlike cloud-centric IoT security frameworks that suffer from high latency our approach enables pervasive intelligence by performing flow-level classification directly on the network gateway.

Our results demonstrate that IoT devices leave stable, discriminative signatures within the first few seconds of activity. Interestingly, we find that extending the observation window does not improve performance, suggesting that identity is most distinct during initial boot-up rather than sustained operation. This allows for rapid, data-efficient enforcement decisions in real-world security applications.

The main contributions of this paper are:
\begin{itemize}
\item We demonstrate that IoT devices can be reliably identified using only the first few seconds of network traffic, without payload inspection.
\item We provide a systematic analysis of observation window duration, showing that early-stage traffic is more informative than longer-term behavior.
\item We design a lightweight, flow-based identification pipeline suitable for deployment on resource-constrained edge devices.
\item We evaluate the approach on a 37 IoT device dataset, achieving up to 99\% accuracy using short observation windows.
\end{itemize}

The remainder of this paper is structured as follows. Section \ref{sec:related} reviews related work on IoT device identification. Section \ref{sec:threat} introduces the threat model. Section \ref{sec:dataset} describes the dataset and experimental setup. Section \ref{sec:methodology} details the methodology, including feature extraction and classification model design. Section \ref{sec:results} reports the experimental results while the implications and limitations of the approach are discussed in Section \ref{sec:discussion}. Finally, Section \ref{sec:conclusion} concludes the paper with concluding remarks and future research directions.

\section{Related Work}
\label{sec:related}
\subsection{IoT Device Identification}
Over the past decade, extensive research has explored IoT device identification through passive network traffic analysis. Prior work primarily employs supervised machine learning models, including Random Forests, Support Vector Machines, and deep neural networks, to classify devices based on statistical traffic characteristics, often achieving high accuracy given sufficiently large labeled datasets \cite{ref1,ref2,ref5,ref6,ref7,ref9,palmese2023designing}. To reduce labeling requirements, unsupervised and semi-supervised approaches such as clustering-based fingerprinting and representation learning using encoder–decoder architectures have also been proposed \cite{ref3, ref18}, though these methods typically lack explicit device-level labeling for enforcement purposes. Feature representations vary across studies, ranging from low-level flow statistics (e.g., packet counts, byte volumes, inter-arrival times) to higher-level contextual attributes such as DNS queries, domains, and port information \cite{ref4,ref16,ref17}. While prior work demonstrates strong performance using long-term traffic aggregation and complex feature pipelines, identification performance is highly dependent on deployment context, device population, and feature design \cite{ref7,ref10}.

\subsection{DNS-based Traffic Analysis}
DNS and DNS-over-HTTPS (DoH) traffic provide valuable signals for network security analysis due to their structured and predictable communication patterns. Even when encrypted, observable metadata such as timing, packet sizes, and flow duration can support classification tasks \cite{ref4}. Prior work has leveraged DNS behavior for anomaly detection and IoT device identification, particularly during device onboarding when DNS resolution is among the first actions performed after network attachment \cite{ref8,ref11}. While many studies aggregate traffic over extended periods to maximise accuracy, recent findings suggest that short observation windows, sometimes as little as one second, can already yield strong identification performance \cite{ref15}. These results motivate further investigation into early-stage identification using limited traffic observations.

\subsection{Data Privacy and Leakage}
Prior research has demonstrated that passive observation of IoT network traffic can reveal sensitive information about devices and user behaviour, even when payloads are encrypted \cite{ref12,ref13,ref14}. Such metadata may expose device presence, type, and usage patterns, highlighting the privacy implications of traffic-based analysis. These findings underscore the dual-use nature of identification techniques: while they support legitimate security and inventory management, they may also enable surveillance if misused. This reinforces the importance of data minimisation and controlled deployment when designing IoT identification systems.

\subsection{Present Study}
In contrast to prior work that relies on long-term traffic aggregation, domain-based fingerprinting, or complex multi-stage pipelines, this study focuses on early-stage IoT device identification using short observation windows. We systematically evaluate how identification accuracy varies as a function of traffic collection duration and demonstrate that high accuracy can be achieved using only the earliest observable network behaviour. Our approach deliberately emphasises low-level statistical flow features, such as timing, volume, and directional characteristics, rather than high-level semantic attributes (e.g., domain names or payload inspection), thereby reducing dependence on infrastructure-specific artefacts.

Furthermore, the system is designed around a modular architecture layering principles, enabling device-agnostic feature extraction and facilitating the straightforward integration of new device classes without modification to the core pipeline. By adopting a defensive network-operator perspective and restricting analysis to lightweight flow-level features, our work complements existing identification and privacy research while emphasizing practicality, extensibility, and early enforcement capabilities.

\section{Threat Model}
\label{sec:threat}
We consider a network environment in which a trusted network operator seeks to identify IoT devices connected to the network using passive traffic observations. The operator has visibility into network metadata at a gateway or monitoring point, such as DNS queries and flow-level statistics, but does not inspect packet payloads or actively interact with devices (as shown in Figure 1).

The objective of the operator is to verify the identity of connected devices in order to support security monitoring, access control, and policy enforcement. We assume that devices are not actively attempting to evade identification and that the network infrastructure itself is trusted.

While similar traffic analysis techniques could potentially be abused by a malicious observer to infer device information, this work focuses on the defensive use of passive identification mechanisms to improve network security and manageability.

\begin{figure}[t]
\centering
\includegraphics[width=\columnwidth]{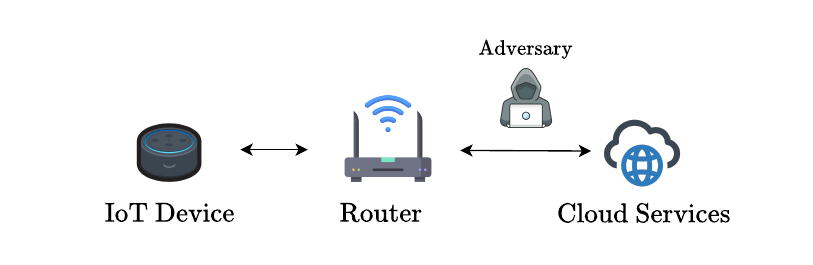}
\caption{Threat model overview showing the trusted network operator monitoring IoT device traffic at the gateway.}
\label{fig:threat_model}
\end{figure}

\section{Dataset and Experimental Setup}
\label{sec:dataset}
\subsection{Dataset Collection}
We collect network traffic from a controlled IoT testbed designed to capture device behaviour immediately following network attachment, as depicted in Figure 2. The testbed consists of multiple IoT devices connected through smart power outlets and a network gateway that passively records all network activity. Device power states are programmatically controlled to enable repeated power-cycling experiments, allowing systematic observation of device initialization behaviour.

Each experiment begins with a device power-on event, followed by continuous traffic capture during the subsequent startup phase. This approach is intended to elicit device-specific communication patterns, such as DNS resolution and initial cloud interactions, which are known to occur shortly after network attachment. Experiments are repeated across multiple days and time periods to capture variability in device behaviour while maintaining consistent experimental conditions.

Our data collection methodology is inspired by prior IoT measurement studies that employ power-cycling and passive traffic monitoring to capture early-stage device communication behaviour \cite{ref9}. While the overall experimental design follows established best practices, the collected data is used here specifically to study early device identification based on limited observation windows.

\subsection{Experimental Setup}
From the collected traffic traces, sessions corresponding to individual device startup events are extracted and processed independently. Each session represents a single device power cycle and includes only traffic observed within a predefined time window following power-on. Flow-level features are computed for each session as described in Section~\ref{sec:methodology}.

To evaluate the feasibility of early identification, we vary the duration of the observation window and assess classification performance as a function of available traffic. Sessions are split into training and test sets at the session level to prevent information leakage across experiments. All model training and evaluation is performed using these session-level representations.

\begin{figure}[t]
\centering
\includegraphics[width=0.85\columnwidth]{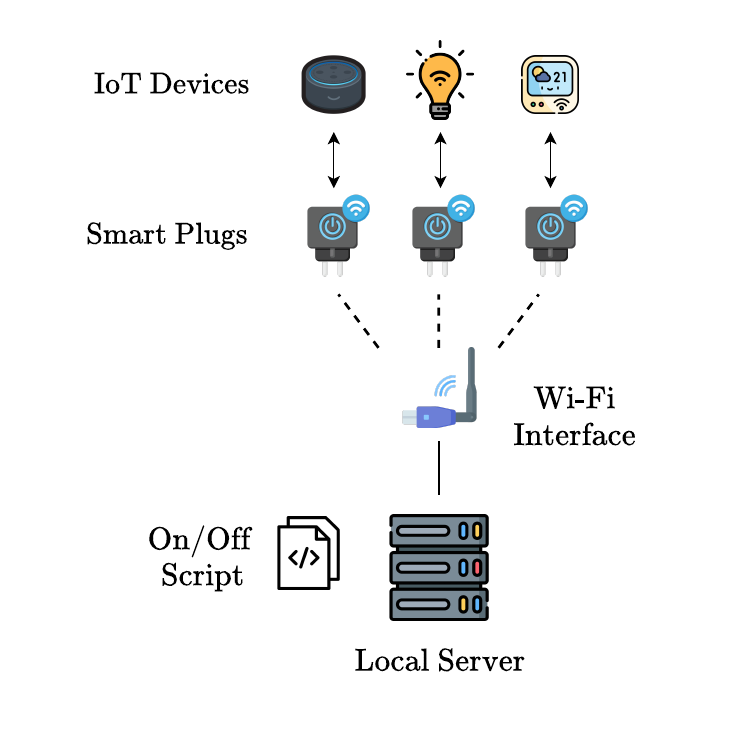}
\caption{Experimental testbed showing IoT devices connected through smart power outlets with passive traffic monitoring at the gateway.}
\label{fig:testbed}
\end{figure}

\section{Methodology}
\label{sec:methodology}

\subsection{Feature Justification}
Table~\ref{tab:features} lists the 40 flow-level features extracted per flow. Features span five broad categories: volume (packet/byte counts and rates), timing (inter-arrival time statistics), TCP flag counts, payload characteristics (entropy and non-zero byte fraction), and categorical metadata (protocol, port bucket, destination type). Payload entropy is computed over the first 512 bytes to capture encryption and encoding patterns without requiring full payload inspection. Port numbers are bucketed into well-known categories rather than used raw, to improve generalization across deployments.

\begin{table*}[t]
\centering
\caption{Flow-level features extracted per flow and their validation status.
  Removal reasons: \textbf{L} = linear combination of other features;
  \textbf{C} = high pairwise correlation ($r \geq 0.9$);
  \textbf{V} = low variance; \textbf{D} = overlapping derived feature.}
\label{tab:features}
\renewcommand{\arraystretch}{0.95}
\setlength{\tabcolsep}{4pt}
\footnotesize
\begin{tabular}{p{4.8cm} p{4.8cm} p{1.6cm} c}
\hline
\textbf{Feature} & \textbf{Meaning} & \textbf{Units} & \textbf{Removed} \\
\hline
\hline
\texttt{dur} & Flow duration & s & --- \\
\hline
\texttt{pkts\_fwd}, \texttt{pkts\_bwd} & Packet counts per direction & count & --- \\
\texttt{pkts\_tot} & Total packet count & count & L \\
\hline
\texttt{bytes\_fwd}, \texttt{bytes\_bwd} & Byte counts per direction & bytes & --- \\
\texttt{bytes\_tot} & Total byte count & bytes & L \\
\hline
{\scriptsize\texttt{pktlen\_fwd\_mean}, \texttt{pktlen\_fwd\_std}} & Forward packet length mean \& std & bytes & --- \\
{\scriptsize\texttt{pktlen\_fwd\_min}, \texttt{pktlen\_fwd\_max}} & Forward packet length min \& max & bytes & C \\
\hline
{\scriptsize\texttt{pktlen\_bwd\_mean}, \texttt{pktlen\_bwd\_std}} & Backward packet length mean \& std & bytes & --- \\
{\scriptsize\texttt{pktlen\_bwd\_min}, \texttt{pktlen\_bwd\_max}} & Backwardd packet length min \& max & bytes & C \\
\hline
{\scriptsize\texttt{iat\_fwd\_mean}, \texttt{iat\_fwd\_std}} & Forward inter-arrival time mean \& std & s & --- \\
{\scriptsize\texttt{iat\_fwd\_min}, \texttt{iat\_fwd\_max}} & Forward inter-arrival time min \& max & s & C \\
\hline
{\scriptsize\texttt{iat\_bwd\_mean}, \texttt{iat\_bwd\_std}} & Backward inter-arrival time mean \& std & s & --- \\
{\scriptsize\texttt{iat\_bwd\_min}, \texttt{iat\_bwd\_max}} & Backward inter-arrival time min \& max & s & C \\
\hline
{\scriptsize\texttt{iat\_tot\_mean}, \texttt{iat\_tot\_std}} & Total inter-arrival time mean \& std & s & --- \\
\hline
\texttt{pps}, \texttt{bps} & Packet \& byte rate & pkt/s, B/s & --- \\
\hline
\texttt{down\_up\_pkt\_ratio} & Down/up packet ratio & --- & --- \\
\texttt{down\_up\_byte\_ratio} & Down/up byte ratio & --- & D \\
\hline
{\scriptsize\texttt{syn\_cnt}, \texttt{ack\_cnt}, \texttt{fin\_cnt}, \texttt{rst\_cnt}} & TCP flags$^{\dagger}$ & count & V \\
{\scriptsize\texttt{psh\_cnt}, \texttt{urg\_cnt}, \texttt{ece\_cnt}, \texttt{cwr\_cnt}} & TCP flags$^{\dagger}$ & count & V \\
\hline
{\scriptsize\texttt{payload\_entropy\_fwd}, \texttt{payload\_entropy\_bwd}} & Payload Shannon entropy per direction & bits/byte & --- \\
\hline
{\scriptsize\texttt{payload\_nonzero\_frac\_fwd}, \texttt{payload\_nonzero\_frac\_bwd}} & Non-zero payload fraction per direction & $[0,1]$ & --- \\
\hline
\texttt{has\_fwd}, \texttt{has\_bwd} & Traffic presence indicators & binary & --- \\
\hline
\texttt{proto} & Transport-layer protocol & categ. & --- \\
\hline
\texttt{sport\_bucket}, \texttt{dport\_bucket} & Port category & categ. & --- \\
\hline
\texttt{internal\_dst} & Destination is RFC1918 & binary & --- \\
\hline
\multicolumn{4}{l}{$^{\dagger}$SYN=Synchronise, ACK=Acknowledgement, FIN=Finish, RST=Reset,} \\
\multicolumn{4}{l}{\phantom{$^{\dagger}$}PSH=Push, URG=Urgent, ECE=ECN-Echo, CWR=Congestion Window Reduced} \\
\end{tabular}
\end{table*}

\subsection{Feature Validation}
In the next steps of building the model, we have identified features to be removed. The decisions have been made by checking if features are part of the following criteria: redundant linear combinations, highly correlated features, low variance features and overlapping features.

\textbf{Redundant linear combinations.} The only features that were removed due to them being linear combinations were packets total and bytes total, as they are the combinations of \texttt{packets\_fwd} and \texttt{packets\_bwd}, and \texttt{bytes\_fwd} and \texttt{bytes\_bwd}, respectively.

\textbf{Highly correlated features.} Linear correlation coefficients have been computed for each pair of features, and those that had a correlation coefficient of at least 0.9 with any other feature have been dropped. Practically speaking, in such analysis, the removed features were the forwrd and backward minimum and maximum values of packet length and inter-arrival times. The practical understanding could be such that the means and standard deviation presented enough information, and the addition of maximal and minimal values added no valuable insights.

\textbf{Low Variance Features.} After testing the variances in the values of features, the removed features were the TCP flags.

\textbf{Overlapping Derived Features.} This is a category of features that can be derived from remaining features; however, they do not exhibit strong linear correlation with any other features. The only feature that has been removed for that reason was \texttt{down\_up\_byte\_ratio} since it is the inverse of the \texttt{down\_up\_pkt\_ratio}.

\subsection{Data Preprocessing}
After the selection of only the relevant features, the next steps involve cleaning the data in the form of removing null values. Then, the data is split between the sessions to prevent occurrences of accuracy degradation due to data leakage---namely, when data coming from a particular session exists both in the training and testing datasets, overfitting is very likely to occur. After doing so the data is split and scaled. Finally, the testing data is being balanced by oversampling.

\subsection{Model}
Instead of a single multiclass classifier, we adopt a one-vs-rest architecture composed of binary Random Forest classifiers, where each model is trained to distinguish a specific device from all others. This design allows incremental addition of new devices without retraining the entire model, which is particularly desirable in dynamic IoT environments where new devices are continuously introduced.
However, this approach introduces a gradual imbalance in the training data as the number of devices increases, which we discuss further in Section VII.
While a multiclass formulation would provide a more compact representation, the one-vs-rest design offers greater flexibility and supports modular deployment at the network edge, allowing for on-device inference in consumer-grade Wi-Fi routers

\textbf{Model Architecture.} We employ a Random Forest classifier as the primary learning model due to its robustness to feature scaling, ability to model non-linear decision boundaries, and strong performance on tabular network traffic data. Random Forests have been widely used in network security applications and provide a favorable trade-off between interpretability and classification accuracy.

Unless otherwise stated, the base model configuration consists of 100 decision trees, with no explicit limit on tree depth, allowing individual estimators to grow until pure leaf nodes or minimum sample constraints are reached. This configuration enables the model to capture complex interactions between traffic features while relying on ensemble averaging to mitigate overfitting. A fixed random seed is used to ensure reproducibility of all experiments.

\textbf{Hyperparameters.} Hyperparameter optimization is conducted using randomised search with cross-validation on the training set. Rather than exhaustively searching the hyperparameter space, we adopt randomised sampling to efficiently explore a broad range of configurations while limiting computational overhead. The following parameters are tuned: number of trees, maximum tree depth, minimum samples required for node splitting, minimum samples per leaf, and the number of features considered at each split. Parameter ranges are selected based on common best practices in the literature and preliminary experiments.

The final model configuration is selected based on mean cross-validation accuracy and subsequently evaluated on a held-out test set that remains unseen during training and tuning. This evaluation protocol ensures that reported results reflect generalization performance rather than optimization artefacts.

\subsection{Data Storage and Deployment}
Since this problem requires that on some occasions the model should be trained on a separate machine to which it is deployed, whereas other times it has to be trained on the same device it is deployed, a cache for intermediate session data has been implemented to speed up the training process in the instances where a different device is used for training to the one the model is deployed to. Namely, the preprocessed data is written to individual parquet files, where each file contains data in one session until a specific number of seconds has passed since the session begun. Also, the sessions metadata is stored in a local key-value store. This allows for data to be quickly retrieved and provide additional insight into the value of different timeframes of collection.

\section{Results}
\label{sec:results}
\subsection{Overall Performance}
The proposed model achieves a test accuracy of approximately 97\%, indicating strong generalisation performance across unseen sessions. Training accuracy approaches 100\%, reflecting high separability of device-specific traffic patterns in the selected feature space. The relatively small gap between training and test accuracy suggests limited overfitting.

\begin{figure}[t]
\centering
\includegraphics[width=\columnwidth]{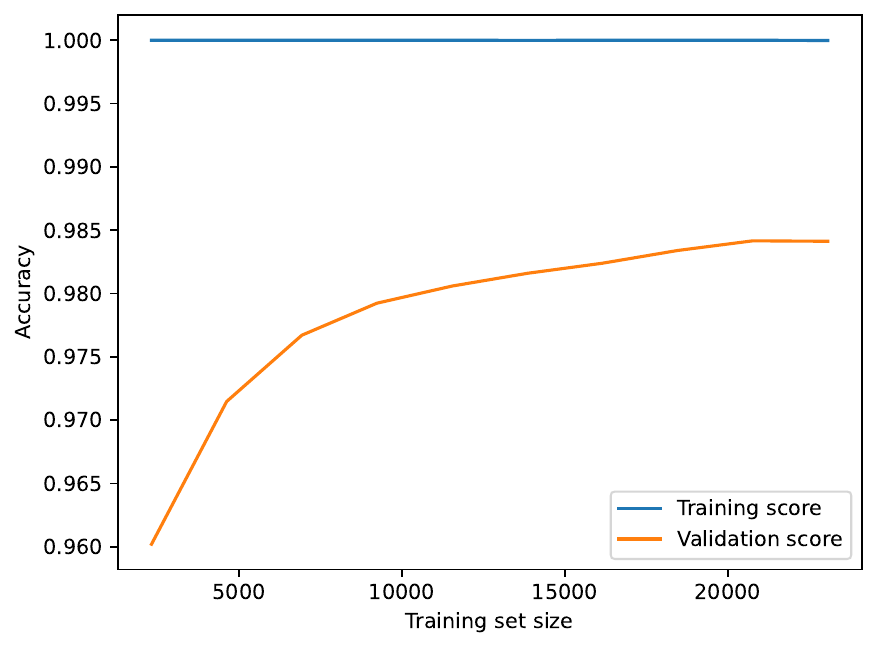}
\caption{Learning curve with train/test accuracy at different training set sizes.}
\label{fig:learning_curve}
\end{figure}

Figure~\ref{fig:learning_curve} illustrates the learning curve of the Random Forest classifier as a function of training set size. The model reaches high accuracy with a relatively small amount of training data, demonstrating that the extracted features are highly informative for device identification. As the training set size increases, test accuracy exhibits a slight downward trend, reflecting increased behavioural variability across sessions rather than degradation in model capacity. Importantly, the gap between training and test accuracy remains stable, indicating controlled generalisation behaviour.

The observed learning behaviour suggests that early-stage network traffic contains consistent, device-specific signatures that can be learned efficiently. The absence of significant performance gains with additional training data implies that the selected feature set captures the dominant discriminative characteristics of the devices under study. This property is particularly desirable for practical deployments, where labeled data may be limited.

Table~\ref{tab:binary_model_results} presents detailed classification performance metrics for each device in the testbed. The majority of devices achieve perfect or near-perfect test accuracy, with only a small number exhibiting classification errors. These results demonstrate the effectiveness of the feature set and model architecture for discriminating between diverse IoT devices.

\begin{table}[t]
\centering
\caption{Summary statistics of binary classification performance across 37 IoT device models (Train Accuracy = 1.000 for all devices)}
\label{tab:binary_model_results}
\renewcommand{\arraystretch}{1.3}
\setlength{\tabcolsep}{6pt}
\small
\begin{tabular}{lcccc}
\hline
\textbf{Metric} & \textbf{Mean} & \textbf{Std} & \textbf{Min} & \textbf{Max} \\
\hline
\hline
Test Accuracy  & 0.987 & 0.029 & 0.860 & 1.000 \\
Precision      & 0.988 & 0.033 & 0.840 & 1.000 \\
Recall         & 0.987 & 0.028 & 0.900 & 1.000 \\
F1-Score       & 0.987 & 0.029 & 0.850 & 1.000 \\
Support        & 315   & 226   & 14    & 913   \\
\hline
\end{tabular}
\end{table}

Selected confusion matrices for representative devices are shown in Table~\ref{tab:confusion_matrix}. Perfect classification is observed for the majority of devices, while occasional misclassifications occur primarily between devices from the same vendor family or with similar cloud service dependencies.

\begin{table}[t]
\centering
\caption{Error analysis across all 37 IoT device classifiers}
\label{tab:confusion_matrix}
\renewcommand{\arraystretch}{1.3}
\begin{tabular}{lcccc}
\hline
\textbf{Error Type} & \textbf{Min} & \textbf{Max} & \textbf{Mean} & \textbf{Devices with 0 errors} \\
\hline
\hline
False Positives & 0 & 17 & 0.41 & 29/37 (78.4\%) \\
False Negatives & 0 & 6 & 0.19 & 32/37 (86.5\%) \\
Total Errors & 0 & 17 & 0.59 & 20/37 (54.1\%) \\
\hline
\end{tabular}
\end{table}

\subsection{Minimum Collection Window}
To evaluate the minimum observation time required for accurate identification, we trained and evaluated the model using incremental traffic windows ranging from 10 to 105 seconds. Across all evaluated intervals, identification accuracy remains consistently high, ranging between 98\% and 99\%. Notably, extending the observation window beyond the initial seconds does not yield consistent performance improvements across devices, and the full window exhibits slightly lower accuracy.
These results indicate that the most discriminative device-specific traffic patterns occur shortly after network attachment. Early communication phases, such as DNS resolution and initial cloud interactions, appear sufficient for reliable identification. Additional traffic collected at later stages introduces behavioural variability without providing new identity-relevant information, leading to a marginal reduction in accuracy.

Table~\ref{tab:time_windows_summary} provides summary statistics. The 30-second window achieves the highest average accuracy (99.4\%) with the greatest proportion of devices achieving perfect classification (75.7\%).

\begin{table}[t]
\begin{center}
\caption{Summary statistics of test accuracy across time windows}
\label{tab:time_windows_summary}
\renewcommand{\arraystretch}{1.3}
\begin{tabular}{lccc}
\hline
\textbf{Time Window} & \textbf{Mean Acc.} & \textbf{Std. Dev.} & \textbf{Perfect (100\%)} \\
\hline
\hline
10 seconds & 98.7\% & 2.2\% & 20/37 (54.1\%) \\
20 seconds & 98.2\% & 3.7\% & 21/37 (56.8\%) \\
\textbf{30 seconds} & \textbf{99.4\%} & \textbf{1.5\%} &\textbf{ 28/37 (75.7\%)} \\
45 seconds & 98.9\% & 1.6\% & 25/37 (67.6\%) \\
60 seconds & 98.4\% & 2.7\% & 23/37 (62.2\%) \\
75 seconds & 99.2\% & 1.7\% & 26/37 (70.3\%) \\
90 seconds & 98.9\% & 1.9\% & 25/37 (67.6\%) \\
105 seconds & 98.9\% & 1.9\% & 25/37 (67.6\%) \\
\hline
\textbf{Overall} & \textbf{98.8\%} & \textbf{2.2\%} & -- \\
\hline
\end{tabular}
\end{center}
\end{table}

\subsection{Error Analysis}
Misclassifications are primarily observed between devices from the same vendor family or with similar backend service dependencies, suggesting that shared infrastructure can partially obscure device-specific signatures. Nevertheless, these errors remain infrequent and do not significantly impact overall identification accuracy.

\section{Discussion}
\label{sec:discussion}
\subsection{Implications of Early IoT Identification}
The results demonstrate that IoT devices exhibit stable and distinctive behaviour immediately after network attachment. High identification accuracy is achievable within short observation windows, indicating that early-stage communication—such as DNS resolution, authentication, and cloud synchronisation—contains strong device-specific signatures. Extending the observation window does not improve performance.

The learning behaviour further indicates that these signatures are consistent across sessions and can be captured with relatively modest training data. From a deployment perspective, this supports rapid enforcement decisions at network gateways without prolonged monitoring or intrusive inspection techniques. Early-stage identification is therefore not only sufficient but preferable for practical security applications.

\subsection{Misclassification Patterns and Limitations}
Misclassifications occur primarily between devices from the same vendor or those sharing backend infrastructure, highlighting the influence of common cloud services on traffic behaviour. While infrequent, such cases suggest an inherent limitation of purely traffic-based identification when device implementations are closely related.

The evaluation is conducted in a controlled environment with a finite device set and firmware versions. Behavioural changes due to software updates, infrastructure modifications, or adversarial traffic obfuscation may affect long-term performance. Future work should therefore evaluate robustness across more diverse deployments and consider resilience against evasion techniques.

\section{Conclusion}
\label{sec:conclusion}
This paper demonstrates that IoT devices can be accurately identified using passive observation of early-stage network traffic. Through systematic evaluation across multiple time windows, we show that device-specific signatures emerge within the first seconds of communication, enabling rapid and data-efficient identification. Our results indicate that extending observation windows provides limited additional benefit, suggesting that early-stage behavior captures the most discriminative characteristics of device identity. By relying solely on flow-level metadata, the proposed approach supports privacy-preserving deployment at network gateways without requiring payload inspection or active probing. These findings highlight the potential of edge-based intelligence for real-time IoT security applications, including device inventory, access control, and anomaly detection.\\
Future work will focus on extending the evaluation to a larger and more diverse device population, including multiple firmware versions and deployment environments. Additional research is needed to assess robustness against encrypted DNS, traffic shaping, and intentional evasion strategies. Integrating the proposed approach with complementary identification signals may further improve resilience in adversarial settings.

\end{document}